\documentclass[12pt,preprint]{aastex}

\newcommand\simlt{\lower.5ex\hbox{$\; \buildrel < \over \sim \;$}}
\newcommand\simgt{\lower.5ex\hbox{$\; \buildrel > \over \sim \;$}}

\begin{document}
\title{Polarization of GRB via Scattering off a Relativistic Sheath}
\author{ David Eichler\altaffilmark{1} and Amir Levinson\altaffilmark{2}}
\altaffiltext{1}{Physics Department, Ben-Gurion University,
Beer-Sheva 84105, Israel; eichler@bgumail.bgu.ac.il}
\altaffiltext{2}{School of Physics \& Astronomy, Tel Aviv
University, Tel Aviv 69978, Israel; Levinson@wise.tau.ac.il}
\begin{abstract}
GRB fireballs could be
ensheathed by a relativistic, baryon-enriched outflow  that would
in selected directions scatter a polarized beam of  gamma rays.
The strong polarization  recently announced  by RHESSI is
attainable by the scattering mechanism, though not most of the
time. The scattering mechanism, when acting in addition to the
synchrotron mechanism,  could circumvent some of the
predictions implied by a purely synchrotron origin for the polarization.
The time profile of the scattered component could differ from that of the
primary gamma rays.
\end{abstract}


\section{Introduction}
  It is easy to imagine that gamma ray burst (GRB) fireballs, which are apparently 
baryon pure, pass through a "birth canal"
ensheathed by an optically thick wall of baryons while the fireballs are 
still within their compact scales. 
 The collimation of the fireball into jets, 
as is now almost universally thought to occur, can be understood 
as the collimating effects of such a sheath (Levinson and Eichler 2000).  
The puzzling problem of baryon purity can be understood if one invokes  an
event horizon that prevents baryons, but not energy, from escaping
along the magnetic field lines that thread it. However, since the
black hole must be fed  matter to power the GRB, matter that is
heated on its way in would also power a baryon rich wind, which would ensheath the
baryon-pure jet (BPJ). The
association of GRB's with supernova suggests that they take place
inside a host star, and then the matter in the host star could in
any event play the role of the sheath.

The BPJ that emanates along
horizon-threading field lines can pick up neutrons that drift in
from the sheath (Eichler and Levinson 1999). It has been recently
argued (Levinson and Eichler 2003) that such neutrons would be
converted by collisions to a roughly equal mixture of neutrons and
protons before penetrating too far into the BPJ. This collisional
avalanche is the basic pick-up mechanism, though it needs to be
jump-started (if by nothing else) by a small number of decaying
neutrons.

An optically thick  but geometrically thin  viscous boundary layer
 develops between the baryon-poor and baryon-rich regions.
 It is dragged along by neutron dominated  viscosity to a Lorentz
factor that can vary extremely rapidly with distance inwards from
the outer wall of baryons. Because the neutrons are coupled by collisions to the
protons, which are trapped on given field lines, the neutrons
cannot freely stream  (fs) into the inner part of the BPJ until
their density drops
 to below a value $n_{fs}$ given by
\begin{equation}
n_{fs}\Gamma_{fs}^{-2}<\sigma v>R/c \le 1.
\end{equation}
 $\Gamma_{fs}$ is determined self-consistently  by the
loading condition $\Gamma_{fs} n_{fs} h \sim F$, where h is the
enthalpy per baryon and F is the flux of the outflow. The value
for $n_{fs}$ so obtained yields a value for  $\Gamma_{fs}$ of
\begin{equation}
\Gamma_{fs}= L_j/(n_{fs}mc^2\pi \theta ^2 r^2ch) = 26
r_{12}^{-1/3} L_{j50}^{1/3}\theta^{-2/3}h^{-1/3} \label{gamma}
\end{equation}
where $r_{12} =R/10^{12}cm, L_{50}=L/10^{50}ergs^{-1}$ and h is
the specific enthalpy in units of the proton rest energy. The
quantity $\Gamma_{fs}$ so derived represents a plausible estimate
of the Lorentz factor of the photosphere, $\Gamma_{ph}$, though
not completely reliable. Photons have a somewhat shorter mean free
path than relativistic neutrons so their free streaming boundaries
are not identical. On the other hand, radiation drag can slow down
the matter to Lorentz factors less than the estimate for
$\Gamma_{fs}$.

The qualitative picture that emerges is that a sharp transverse
density contrast is established by the pick-up of neutrons that
leak in from the walls of the BPJ. The innermost optical depth of
the baryon-enriched viscous transition layer (i.e. the inner
photosphere) can be moving relativistically $\Gamma_{ph} \gg 1$.
This is what we hypothesize. Photons that are generated in the BPJ interior can 
scatter off this relativistic sheath and be scattered by an  angle 
$1/\Gamma_{ph}$ from the velocity vector of the sheath. If the angle $1/\Gamma_{ph}$
is larger than the opening angle $\theta_v$ of the velocity cone of the sheath, then the solid 
angle over which the GRB is viewable is about $\pi 1/\Gamma_{ph}^2$. Otherwise, it is 
defined by an annulus  $\theta_o$ with thickness $1/\Gamma_{ph}$. 

Below we consider the consequences for polarization of a significant fraction of
the photons emitted in the BPJ scattering of the baryon-enriched
sheath. We  make the reasonable assumption that, at
sufficiently large radius, the photon mean free path can be
large compared to  the scale of the density scale height. This
follows naturally if there is transverse density stratification:
the inner tenuous high $\Gamma$ regions become transparent at
smaller radii (Eichler and Levinson 2000) than the outer regions,
which are more heavily baryon-enriched. As illustrated in figure 1, 
the radius of the inner photosphere 
is thus an increasing function of transverse distance from the axis. 
A photon emitted in an optically thin region near the axis (at point A)
 could scatter off the photosphere 
at a point (B) that is at larger radius from the source and larger 
cylindrical radius from the axis than 
point A.
Given that the photon mean
free path is large, a   scatterer close to the photosphere sees  a highly anisotropic bath
of photons and this allows for polarization via scattering.

Although we have in our  earlier work  argued  (for other
motivations)for the particular geometry in which the soft photon
source is ensheathed by the scatterer, we note that, for the
purposes of polarization,  the scattering geometry proposed by
Lazzati et. al (2000), in which the  scatterer is within a
surrounding  soft photon bath, would work just as well (though the
implications for time variablility of the polarized component
might be different). We also note that Begelman and Sikora (1987)
many years ago proposed the same mechanism for polarization of
blazar emission, (see also Dermer and Schlickheiser, 1993). Our
polarization calculations, as far as we can tell, agree with
theirs.  Here we also calculate the likelihood that a previously
unidentified source would display a particular polarization for a
homogenous distribution of sources.

\section{Polarization by Scattering}

In the geometry illustrated in figure 1, which is but one of several variations, slight
collimation of the BPJ is portrayed, and the opening angle of the
scattered photons is larger than  the final opening angle of the
BPJ. The collimation could be stronger than indicated in the
figure, and then  most of the photons emitted by the BPJ, no matter how
strongly beamed forward, would be intercepted by the walls before
escaping. Alternatively, the  opening angle of the  scattered photons 
could be less than the opening 
angle of the velocity cone of the photosphere, 
and in this case the scattered photons form an annulus. 
We do not need to assume that most photons are scattered off the sheath, only that 
most photons in the observer's direction are.  

Consider a fluid element A just axisward of the photosphere as in
figure 1. Half of its photons are beamed outward relative to the
local direction of the outflow. They enter the denser viscous
transition layer and are scattered in fluid element B, which is by 
assumption near the photosphere. Assuming
that $\Gamma$ varies very rapidly  through the value of
$\Gamma_{ph}$, then in the frame of B, they are moving  nearly in
the direction of the outflow itself, and scatter off B after a rear approach. 
Alternatively, if they enter
element B from most other directions (as seen in the lab frame),
then in the frame of B they are moving nearly opposite to B's
motion (referred to by Begelman and Sikora [1987] as the head-on
approximation). If they now scatter off an electron in element B,
the polarized emissivity as a function of angle $\psi'_B$ in frame
B satisfies, in the limit of a perfect photon beam,
$j^\prime_{pol}(\psi^\prime_B)\propto 1-\cos^2\psi^{\prime}_B$,
and the total intensity $j^\prime_{tot}(\psi^\prime_B)\propto
1+\cos^2\psi^{\prime}_B$. By the aberration relation of $\cos
\psi^{\prime}$ to $\cos\psi$, they can also be expressed as a
function $\psi$ in the lab frame (assuming a power law spectrum
for the scattered photons)
\begin{equation}
I_{pol}(\psi)\propto\frac{\left[(1-\beta \cos\psi)^2 - (\cos\psi -
\beta)^2)\right]}{\Gamma^k(1-\beta\cos\psi)^{k+2}},
\label{I_pol}
\end{equation}
and
\begin{equation}
I_{tot}(\psi)\propto\frac{\left[(1-\beta \cos\psi)^2 + (\cos\psi -
\beta)^2)\right]}{\Gamma^k(1-\beta\cos\psi)^{k+2}}.
\label{I_tot}
\end{equation}
Here $\cos\psi=\hat{n}\cdot\hat{\beta}$, where $\hat{n}$ and
$\hat{\beta}$ are unit vectors along the line of sight and in the
direction of fluid element B, respectively, and the index $k$
depends on the spectral index of the scattered radiation, and the
kinematics of the scattering region. For a spectral index of
$\alpha$, $k=\alpha+2$ if the center of emissivity is stationary
with respect to the observer (e.g., a continues jet) and  $k=\alpha+3$  
if the center of emissivity moves with velocity $\beta c$  (e.g, a discrete
blob; see Lind and Blandford 1985).  Typically $k$
lies in the range $2.5$ to $3.5$. Below we adopt $k=3$ for
illustration We suppose that all scatterers are moving on a cone
of opening angle $\theta_0$. Choosing a local coordinate frame
such that $\hat{n}=\sin\theta \hat{x}+\cos\theta\hat{z}$, and
$\hat{\beta}=\sin\theta_0\cos\phi
\hat{x}+\sin\theta_0\sin\phi\hat{y}+\cos\theta_0\hat{z}$, and
defining $\hat{t}=
\hat{n}\times\hat{\beta}/|\hat{n}\times\hat{\beta}|$, and
$\hat{b}= \cos\theta \hat{x}-\sin\theta\hat{z}$ ($\hat{t}$ is a unit vector 
normal to the velocity vector as projected on the plane of the sky, and
$\hat{b}$ defines the 
axis of symmetry on the plane of the sky), the Stokes
parameters for a given sight line integrated over all scatterers
directions can be expressed as

\begin{equation}
Q(\theta)=\int{I_{pol}(\psi)\cos(2\chi)d\phi},
\label{I_polav}
\end{equation}
and
\begin{equation}
I(\theta)=\int{I_{tot}(\psi)d\phi},
\label{I_totav}
\end{equation}
where $\cos\chi=\hat{t}\cdot\hat{b}$.
(The Stock parameter $U$ vanishes for the above choice of coordinate system.)
The corresponding polarization degree is given by
\begin{equation}
P(\theta)= |Q(\theta)|/I(\theta).
\label{pol}
\end{equation}
Eqs. (\ref{I_polav}) and (\ref{I_totav}) have been integrated numerically.
Fig. 2 shows the resultant polarization degree versus
$\Gamma(\sin\theta_0-\sin\theta)$ for different values of
$\Gamma\theta_0$.  As seen, large polarization can be observed in
the case of viewing from outside a narrow velocity cone  (that is,
$\Gamma\theta_0<<1$) for viewing angles $\theta\sim 1/\Gamma$. For
this case, the polarization is always in the same direction, and,
if the velocity vector of the scattering material were to wobble
or be spread by less than $\theta$, there would still be a net
polarization.

If, however, most GRB's are viewed from an annulus at angular
inset $\delta \theta<<\theta_0$ from a cone of opening angle
$\theta_{0}$,  then the observed polarization would typically be
smaller.  The relative probabilities are plotted in figure 3 both
for  a narrow beam with spread $1/\Gamma_{ph}$ and a thin inset
annulus for homogeneous distribution. In the case of viewing from
within the thin inset annulus, the kink in the curves corresponds
to a 90 degree change in the polarization vector.

If the  primary emitter does not have large $\Gamma$ relative to
the scattering material, then the photons can enter the latter
neither from a parallel, overtaking direction (i.e. from behind)
nor from the anti-parallel direction, but, rather, offset by a
small angle from the parallel direction. In this case, the
direction of total polarization is beamed forward and strong
polarization can take place with less of a sacrifice of intensity.

\section{Discussion}

In independent earlier work (Nakamura 1998, Eichler and Levinson
1999) it was noted that GRB980425, thought to be associated with
supernova 1998bw, had an unusual light curve. The lack of multiple
peaks and the relatively soft spectrum were suggestive of a
scattered component, where the spikiness of the typical GRB light
curve could be washed out by multiple-path scattering into the
observer's line of sight. The extremely small gamma ray luminosity
implied by the association with the relatively nearby SN1998bw is
also consistent with the hypothesis that it was scattered off some
sort of baryonic sheath. Finally, the unusually large ejection
velocity observed in the supernova remnant may suggest that it was
"dragged" by the GRB fireball, possibly by the interaction of high
energy neutrinos from the BPJ with the surrounding baryonic
material.

GRB021206 on the other hand, was bright, had a typical light curve
containing several peaks, and substantial emission above 1 MeV.
There is no particular {\it a priori} reason to suspect scattering
off slowly moving material. However, scattering off a {\it
relativistic sheath} could preserve hardness and spikiness; the
photon energy in the frame of the scatterer is still in the
Thompson limit and the relativistic beaming consolidates the area
on the scattering surface that is capable of scattering photons
into the observer's line of sight.

Synchrotron emission in an ordered magnetic field is, for this
particular burst, a reasonable explanation for the high degree of
polarization reported by the original discovery paper (Coburn and
Boggs 2003, hereafter CB). It was also argued  by  CB that
scattering would imply high angular dilution and a small optical
depth, both of which would considerably raise the overall energy
requirement of the (already bright) GRB021206. Nevertheless, we
consider the polarization mechanism in view of the following
points: First, invoking a relativistic flow for the scattering
material (relative to the emitter) perpendicular to the density
gradient eliminates the two objections of CB. Relativistic beaming
of the scattered radiation allows it to remain collimated. That
the incident photons as seen by the scattering material arrive
from a direction nearly perpendicular to the density gradient
implies that they penetrate the density contours of the scattering
material  at a very glancing angle $\sim 1/\Gamma_{ph}$ and always
interact at very low optical depth along the shortest path out. In
the directions corresponding to strong polarization, i.e. 90
degree scattering - nearly along the density gradient, there is
negligible attenuation by a second scattering despite the high
optical depth to a first scattering. That we have ignored the
opacity encountered by the scattered radiation on its way out of
the scattering material suggests that we may have in fact {\it
underestimated} the polarization. We have also neglected multiple
scattering within the scattering material which, when included,
would {\it lower} the estimated polarization. However,
multiply-scattered photons within a diverging relativistic outflow
would suffer energy losses; the energy would be returned to the
matter outflow. Efficiency considerations suggest that the inner
photospheric material suffers radiation drag, and that it be kept
relativistic by viscous interaction  with the deeper layers.
Energy lost by multiply scattering photons can thus be recovered
and rechanneled into polarized radiation.

Scattering  can be a significant  contributor to polarization
without necessarily being the only mechanism.  The experimental
consequences of such a contribution could be polarization that is
occasionally above the theoretical limit for a pure optically thin synchrotron
origin. This would require that both mechanisms are working in the
same direction of polarization;  for polarization that is parallel
to the axis of symmetry (as projected onto the sky) this would
mean scattering within an angle less than $1/\Gamma$ from the
plane of the scattering surface, and would allow at most a 25
percent increase.

Synchrotron emission from an optically thin emitting region has
been criticized as being a universal emission mechanism for GRB's
on the grounds that it does not for any obvious reason produce
energy peaks consistently at 200 KeV with the observed sharpness
(e.g.  Eichler and Levinson 2000). Even  a monochromatic electron
population would produce synchrotron spectra that for some GRB's
contain more than the observed proportion of low energy photons
(Preece et al. 2002). Even if the frequently observed paucity of
low energy photons could be attributed to the low energy cutoff in
the electrons or to self-absorption, a strong energy dependence in
the polarization would be implied at low energies. This could be
tested. A purely scattering  origin for polarization would, on the
other hand, predict polarization that is frequency independent at
low energies.

Synchrotron   and scattering mechanisms for polarization could
conceivably be distinguished from each other by their time
profiles. A scattered  component of the polarized emission would
have a time profile that was smoothed out on timescales less than
$r/c\Gamma^2$ where r is the characteristic scale at which the
last scattering takes place. Future experiments that can measure
polarization on sub-second timescales would therefore be extremely
valuable.

Generally speaking, scattering, when operating alone as a
polarizing mechanism,  produces somewhat lower polarization than
optically thin synchrotron radiation. It will not compete with the
latter if polarizations of 0.55 are consistently observed.
However, it is capable of producing even stronger polarization
than synchrotron sometimes, so an occasional polarization near 100
percent, if it could be measured with sufficient precision,  would
suggest scattering. In any case, we have argued that, with the
unfolding technology to detect gamma ray polarization,
polarization by scattering may be detectable and distinguishable
from polarization by synchrotron emission.

We thank E. Derishev  for useful discussions. This research was
supported by the Arnow Chair of Astrophysics at Ben Gurion
University and by an Adler Fellowship awarded by the Israel
Science Foundation.

\clearpage


\section{Figure captions}

\figcaption{A sketch of the collimated fireball geometry is
shown. The inner part is the unshocked baryon-poor jet (BPJ),
which is collimated by the baryon rich material enveloping it. At
the interface, a viscous transition layer is established by
neutron leakage which is dragged to relativistic Lorentz factor
 as it is picked-up by the BPJ. Photons scattering off
the inner part of the viscous transition  layer are scattered into
an annulus or cone within $1/\Gamma_{ph}$ of the velocity vector
of the  photosphere, where $\Gamma_{ph}$ is the Lorentz factor of 
the photospheric material.}

\figcaption{ The polarization degree as a function of
$\Gamma(\sin\theta_0-\sin\theta)$ for different values of
$\Gamma\theta_0$ for scattering off a thin hollow cone. 
 Negative values on the horizontal axis
correspond to the region inside the cone of scattering material.
The kinks in the curves correspond to a 90 degree change in the
polarization vector. The polarization in the inner bump of the P
curves (around $\theta=\theta_0$) is parallel to the projection of the
jet axis on the sky.}

\figcaption{The probability  is displayed that photons, if scattered
from material  moving on a velocity cone of opening angle
$\theta_0$, will be observed to have a polarization higher than
$\pi$.  The ankle in the dotted and dot-dashed curves corresponds to the
change in polarization direction. The photons are assumed to
impinge on each scatterer from either the forward or reverse
direction in the scatterer frame. The  sources are assumed to be
uniformly distributed in a Euclidean space; for $<V/V_{max}>$
significantly less than 0.5, the probabilities for high
polarization would be higher. The material is assumed to be
optically thin.}

\newpage
\plotone{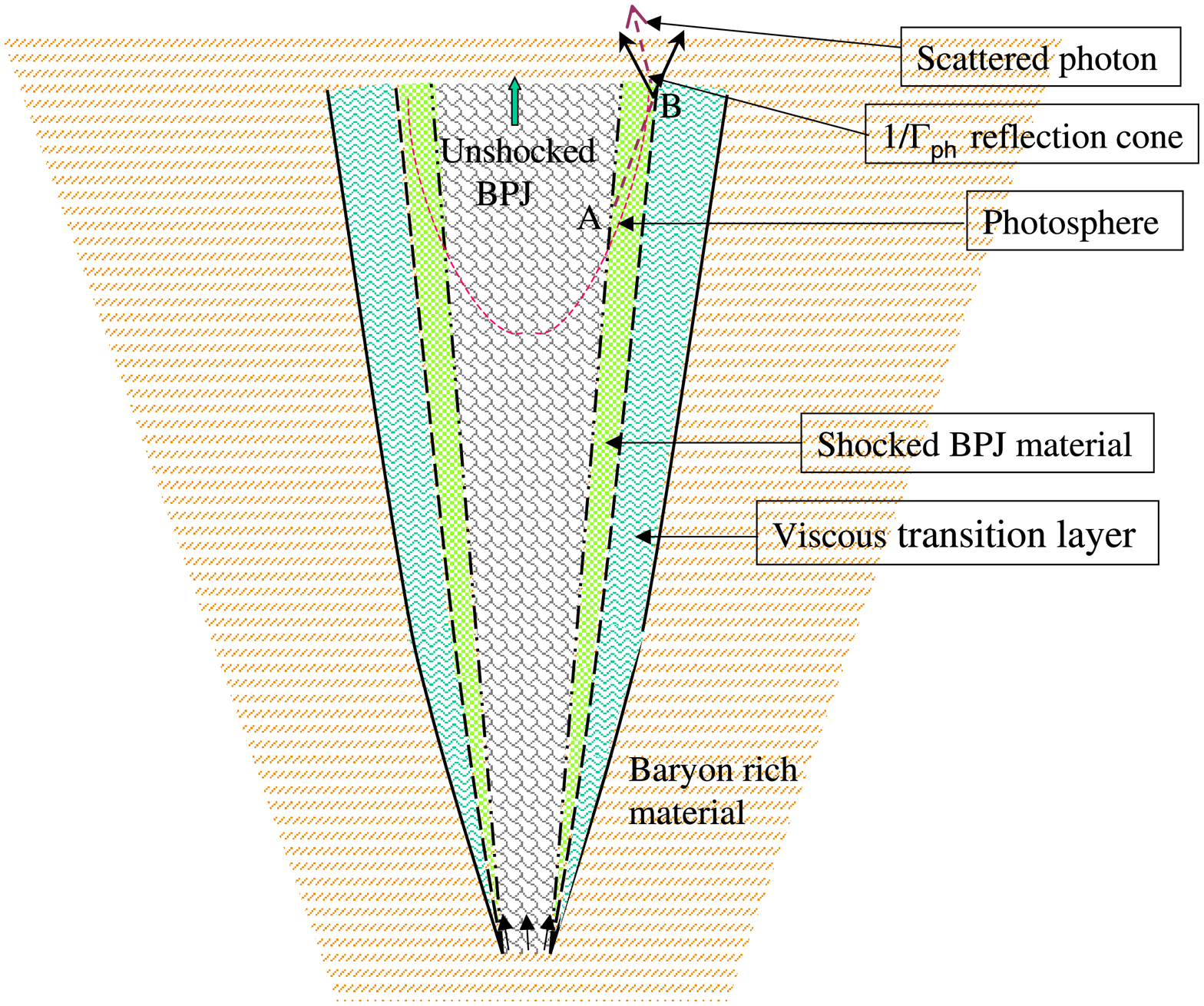}
\newpage
\plotone{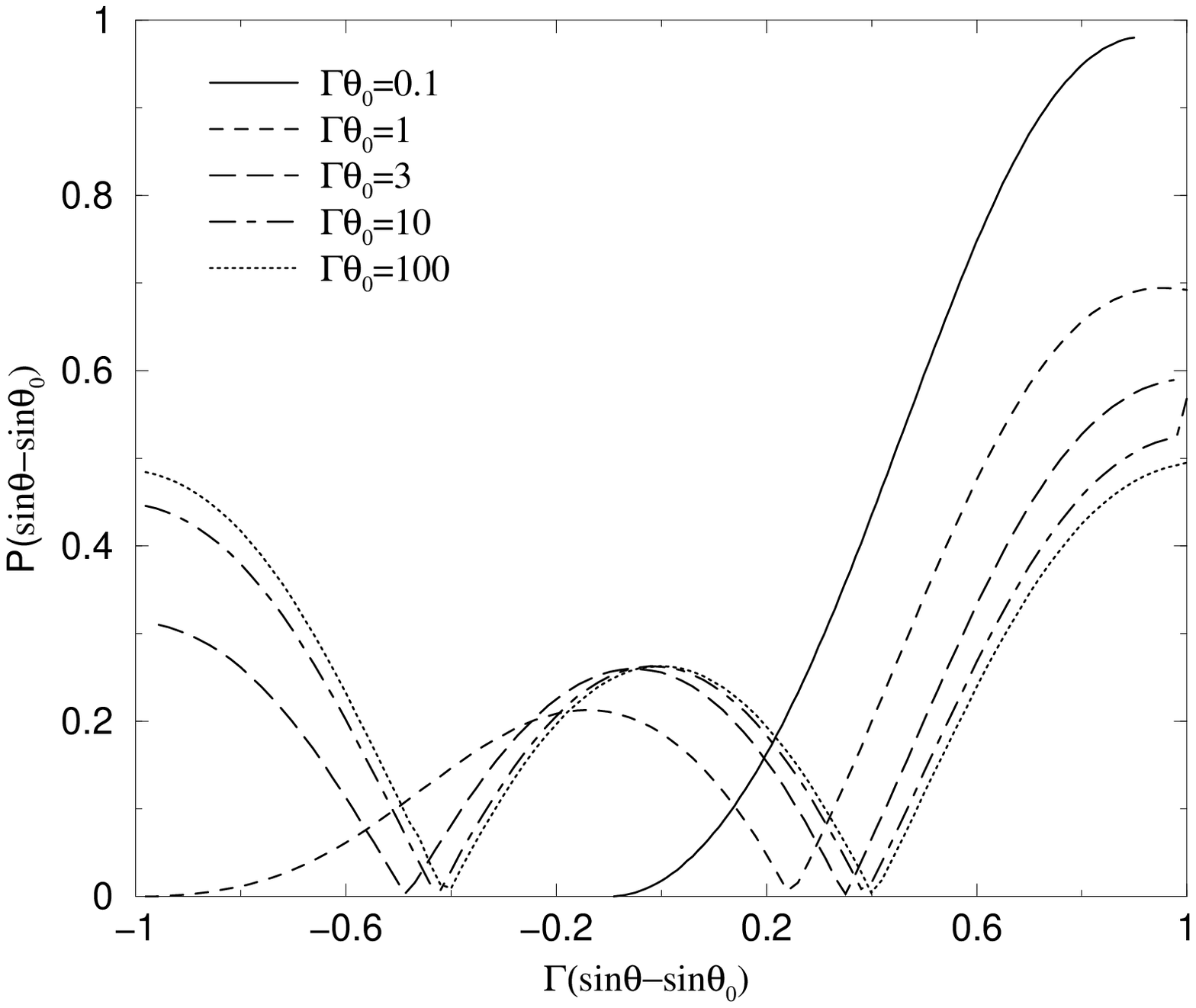}
\newpage
\plotone{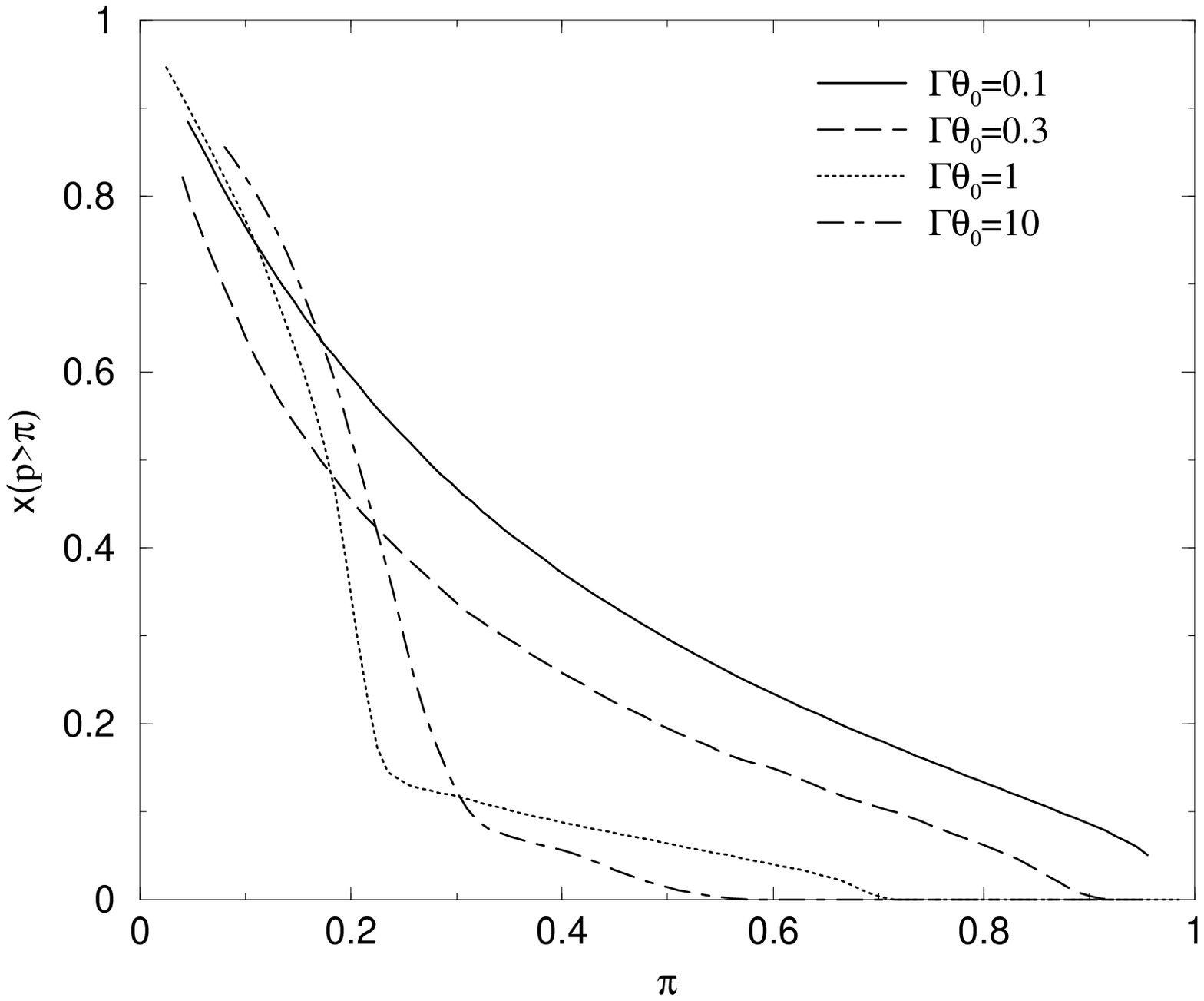}


\begin{thebibliography}{99}
\bibitem[]{348} Begelman, M.C., and Sikora, M. 1987, ApJ 322, 650
\bibitem[]{349} Coburn, W. \& Baggs, S.E. 2003, Nature, 423, 415
\bibitem[]{356} Eichler, D. \& Levinson, A. 1999, ApJ, 521, L117
\bibitem[]{357} Eichler, D. and Levinson, A. 2000, ApJ 529, 146
\bibitem[]{361} Lazzati, D. Ghisselini, G. Celotti A. and Rees,
M.J., 2000 ApJ, 529, L17
\bibitem[]{362} Levinson, A. \& Eichler, D. 1993, ApJ, 418, 386
\bibitem[]{363} Levinson, A. \& Eichler, D. 2003, astro-ph/0302569
\bibitem[]{364_a} Lind, K.R., \& Blandford, R.D. 1985, ApJ 295, 358
\bibitem[]{364_b} Nakamura, T., 1998, Prog. Theor. Phys., 100, 921
\bibitem[]{368} Preece, R.D. et al. 2002, ApJ, 581, 1248
\end{thebibliography}
\end{document}